\documentclass[twocolumn,aps,showpacs,preprintnumbers,amsmath,amssymb, superscriptaddress]{revtex4-1}
\pdfoutput=1

\usepackage{bm}
\usepackage{graphicx}
\usepackage{color}

\begin{document}

\title{Angle-resolved photoemission spectroscopy observation of anomalous electronic states in EuFe$_2$As$_{2-x}$P$_x$}

\author{P. Richard}\email{p.richard@iphy.ac.cn}
\affiliation{Beijing National Laboratory for Condensed Matter Physics, and Institute of Physics, Chinese Academy of Sciences, Beijing 100190, China}
\author{C. Capan}
\affiliation{Department of Physics and Astronomy, University of California, Irvine, California 92697, USA}
\affiliation{Department of Physics, Washington State University, Pullman, WA}
\author{J. Ma}
\affiliation{Beijing National Laboratory for Condensed Matter Physics, and Institute of Physics, Chinese Academy of Sciences, Beijing 100190, China}
\author{P. Zhang}
\affiliation{Beijing National Laboratory for Condensed Matter Physics, and Institute of Physics, Chinese Academy of Sciences, Beijing 100190, China}
\author{N. Xu}
\affiliation{Beijing National Laboratory for Condensed Matter Physics, and Institute of Physics, Chinese Academy of Sciences, Beijing 100190, China}
\affiliation{Paul Scherrer Institut, Swiss Light Source, CH-5232 Villigen PSI, Switzerland}
\author{T. Qian}
\affiliation{Beijing National Laboratory for Condensed Matter Physics, and Institute of Physics, Chinese Academy of Sciences, Beijing 100190, China}
\author{J. D. Denlinger}
\affiliation{Advanced Light Source, Lawrence Berkeley National Laboratory, Berkeley, California 94720, USA}
\author{G.-F. Chen}
\affiliation{Beijing National Laboratory for Condensed Matter Physics, and Institute of Physics, Chinese Academy of Sciences, Beijing 100190, China}
\author{A. S. Sefat}
\affiliation{Materials Science and Technology Division, Oak Ridge National Laboratory, Oak Ridge, Tennessee 37831-6114, USA}
\author{Z. Fisk}
\affiliation{Department of Physics and Astronomy, University of California, Irvine, California 92697, USA}
\author{H. Ding}
\affiliation{Beijing National Laboratory for Condensed Matter Physics, and Institute of Physics, Chinese Academy of Sciences, Beijing 100190, China}

\date{\today}

\begin{abstract}
We used angle-resolved photoemission spectroscopy to investigate the electronic structure of EuFe$_2$As$_2$, EuFe$_2$As$_{1.4}$P$_{0.6}$ and EuFe$_2$P$_2$. We observed doubled core level peaks associated to the pnictide atoms, which are related to a surface state. Nevertheless, strong electronic dispersion along the $c$ axis, especially pronounced in EuFe$_2$P$_2$, is observed for at less one band, thus indicated that the Fe states, albeit probably affected at the surface, do not form pure two-dimensional surface states. However, this latter material shows reduced spectral weight near the Fermi level as compared to EuFe$_2$As$_2$ and EuFe$_2$As$_{1.4}$P$_{0.6}$. An anomalous jump is also found in the electronic states associated with the Eu$^{2+}$ $f$ states in EuFe$_2$P$_2$.  
\end{abstract}

\pacs{74.70.Xa, 74.25.Jb, 79.60.-i}


\maketitle

Although not as extensively studied as the Ba$_{1-x}$K$_x$Fe$_2$As$_2$ and BaFe$_{2-x}$Co$_x$As$_2$ archetype systems of 122-ferropnictides, the EuFe$_2$As$_{2-x}$P$_{x}$ compounds show very unique and exotic properties, which vary significantly upon As$\rightarrow$P isovalent substitution. In addition to a magnetic Fe network, a large moment is observed on the Eu$^{2+}$ ions \cite{Raffius_JPCS54,Z_RenPRL2009,Y_XiaoPRB80,C_Feng_PRB82, Ryan_PRB83}. While the Eu sublattice exhibits A-type antiferromagnetism below $T_N=19$ K in EuFe$_2$As$_2$ \cite{Y_XiaoPRB80,Herrero_MartinPRB80}, a ferromagnetic structure with slightly canted Eu moments aligned along the $c$-axis is observed in EuFe$_2$P$_2$ below $T_N=30$ K \cite{C_Feng_PRB82, Ryan_PRB83}. The Eu atoms also seem to play an essential role in the anomalous compressibility effects observed in EuFe$_2$As$_2$ \cite{UhoyaJPCM22}, and a Eu valence change under pressure has even been reported in superconducting EuFe$_2$As$_{1.4}$P$_{0.6}$ \cite{LL_SunPRB82}.

Especially following the discovery of reentrant superconductivity in EuFe$_2$As$_{1.3}$P$_{0.7}$ coinciding with the ordering of the Eu$^{2+}$ moments \cite{Z_RenPRL2009}, the detail of the interplay between the Eu$^{2+}$ and Fe$^{2+}$ layers, as well as the precise role of the As$\rightarrow$P isovalent substitution for the emergence of superconductivity, became important issues that are still debated and need proper experimental characterizations. With its capacity to resolve the one-particle electronic spectra of materials directly in the momentum space, angle-resolved photoemission spectroscopy (ARPES) is a powerful tool that may be used for such purposes. Indeed, the electronic structures of the EuFe$_2$As$_2$ parent compound \cite{ZhouBo_PRB81,Adhikary_JPCM25,Thirupathaiah_PRB84} and of EuFe$_2$As$_{1.56}$P$_{0.44}$ \cite{Thirupathaiah_PRB84} have been studied by ARPES recently. Although it has been first synthesized \cite{MarchandJSSC24} three decades before the discovery of Fe-based superconductivity in 2008 \cite{Kamihara_JACS2008}, there is unfortunately no ARPES report in literature on the electronic structure of EuFe$_2$P$_2$.

In this paper, we present an ARPES study of the electronic structure of EuFe$_2$As$_{2-x}$P$_{x}$, from EuFe$_2$As$_2$ to EuFe$_2$P$_2$. The photoemission spectra indicate that all these materials have at least 2 inequivalent pnictide sites, which is linked to a surface state possibly resulting from the pnictide-pnictide interactions occurring in short $c$-axis 122 compounds. Nevertheless, we record strong modulations of the electronic structure along the perpendicular momentum ($k_z$) direction, which become more prominent upon As$\rightarrow$P substitution. We also observe an unexplained jump in the energy position of the Eu$^{2+}$ $f$ electrons in EuFe$_2$P$_2$.

Single-crystals of EuFe$_2$As$_2$, EuFe$_2$As$_{1.4}$P$_{0.6}$ and EuFe$_2$P$_2$ were grown using conventional methods described in Refs. \cite{Sefat_PRL2008,LL_SunPRB82,Canfield_PMB1992}. While the typical size of the samples of the first two compounds exceeds $1.5\times 1.5$ mm$^2$, much smaller samples (around $150\times 150$ $\mu$m$^2$) of EuFe$_2$P$_2$ were measured. Most of the ARPES measurements were performed at the PGM and APPLE-PGM beamlines of the Synchrotron Radiation Center (Wisconsin) equipped with a VG-Scienta R4000 analyzer and a SES 200 analyzer, respectively. The energy and angular resolutions for the angle-resolved data were set at 10 - 30 meV and 0.2$^{\textrm{o}}$, respectively. The samples were cleaved \emph{in situ} and measured at 20 K in a vacuum better than $5\times 10^{-11}$ Torr. Additional core level measurements of the EuFe$_2$As$_2$ surface under potassium evaporation were performed at the Merlin beamline of the Advanced Light Source (California). We label the momentum values with respect to the 1 Fe/unit-cell Brillouin zone (BZ), and use $c^{\prime} = c/2$ as the distance between two Fe planes.

\begin{figure}[!t]
\begin{center}
\includegraphics[width=3.4in]{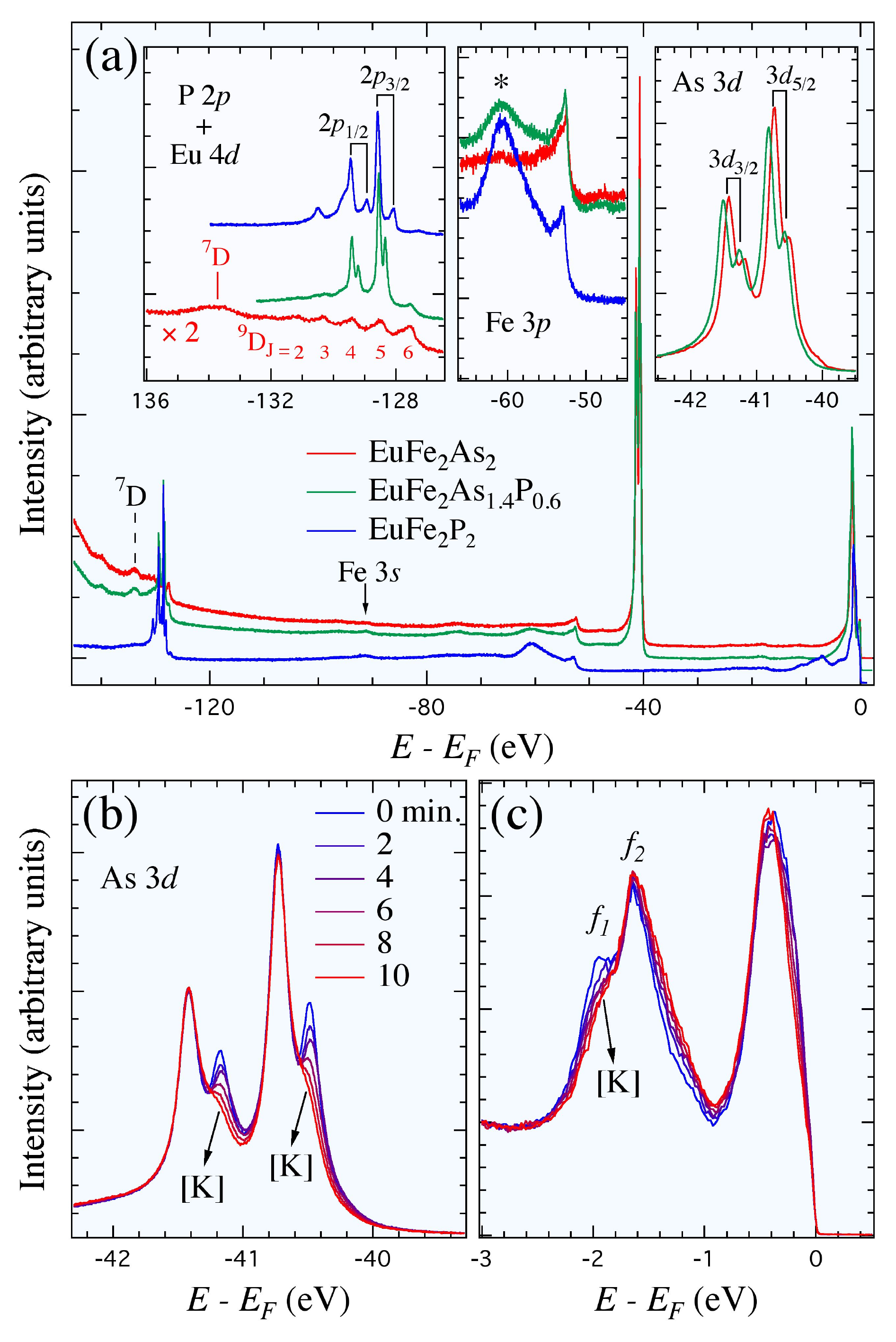}
\end{center}
\caption{\label{core}(Color online). (a) Core level spectra of EuFe$_2$As$_{2-x}$P$_{x}$ recored with 180 eV photons. The spectra have been shifted vertically for a better visualization. The left inset corresponds to the P $2p$ and Eu $4d$ spectral range, where the spectrum of EuFe$_2$As$_2$ has been multiplied by 2. Black lines are guides to the eye for the splitting between the peaks of two sites of P. The middle inset coincides with the Fe $3p$ spectral range. The asterisks refers to 115.3 eV Auger electrons from P. The right inset corresponds to the spectral range of As $3d$ electronic states. Black lines serve as guides to the eye for the separations between the peaks of two As sites. (b) and (c) Evolution of the photoemission spectra as a function of the time of potassium evaporation, for the As $3d$ and the near-$E_F$ spectral ranges, respectively. }
\end{figure}

In Fig. \ref{core}(a), we compare the core level spectra of EuFe$_2$As$_2$, EuFe$_2$As$_{1.4}$P$_{0.6}$ and EuFe$_2$P$_2$, which show signatures of the elemental composition of these materials. We first describe the features observed above 20 eV of binding energy ($E_B$). All spectra contain a very small peak detected around $E_B=91.3$ eV that we assign to the Fe 3$s$ electronic state, as well as a well-defined and more intense peak associated with the Fe 3$p$ electrons, which is observed at a $E_B$ of 52.4 eV, 52.6 eV and 52.9 eV in these three compounds, respectively. As emphasized in the middle inset, a broad bump absent in EuFe$_2$As$_2$ but increasing with P substitution is also observed at a slightly higher $E_B$ (60.7 eV). This bump is detected at the same kinetic energy of 115.3 eV, independently of the incident photon energy, and we thus attribute it to Auger electrons from P. 

The most intense peaks observed in EuFe$_2$As$_2$ and EuFe$_2$As$_{1.4}$P$_{0.6}$ correspond to the As $3d_{3/2}$ and $3d_{5/2}$ states. A zoom, displayed in the right inset of Fig. \ref{core}(a), shows an average shift of 75 meV towards the high $E_B$ in EuFe$_2$As$_{1.4}$P$_{0.6}$ as compared to EuFe$_2$As$_2$. More importantly, we observe that these peaks are doubled, indicating the presence of two inequivalent As sites, suggesting a surface reconstruction affecting directly the As electronic states. When increasing the P content from $x=0$ to $x=0.6$, the splittings between the peaks associated with the two sites increase slightly, from 237 meV to 248 meV, and from 225 meV to 240 meV for the As $3d_{3/2}$ and As $3d_{5/2}$ states, respectively. As expected from their equivalent role in the structure of EuFe$_2$As$_{2-x}$P$_{x}$, double-peak features are also observed for the P $2p_{1/2}$ and P $2p_{1/2}$ electronic states in the P-substituted materials. As shown in the left inset of Fig. \ref{core}(a), the splitting between the two sites increases significantly as the P concentration is raised from $x=0.6$ to $x=2$. Indeed, we record a splitting that increases from 206 meV to 522 meV for the P $2p_{1/2}$ states, and a splitting that increases from 210 meV to 519 meV for the P $2p_{3/2}$ states. It is important to note that while such effect on the pnictide atoms could potentially have a sizable impact on the Fe electronic states in EuFe$_2$As$_{2-x}$P$_{x}$, our previous measurements \cite{Neupane_PRB2011} on a wide doping range of hole-doped Ba$_{1-x}$K$_x$Fe$_2$As$_2$ and electron-doped BaFe$_{2-x}$Co$_x$As$_2$ did not evidence any of these doubled features. 

To determine whether the doubled features observed in these materials are related to a surface state or to an impurity phase, we evaporated successively small amounts of K atoms \emph{in situ} on the cleaved surface of a EuFe$_2$As$_2$ sample and investigated the As $3d$ core levels. As illustrated in Fig. \ref{core}(b), the high-$E_B$ components of the As $3d_{3/2}$ and As $3d_{5/2}$ core levels are barely affected by this process, suggesting that they are related to bulk states. In contrast, the low-$E_B$ components are rapidly suppressed upon K evaporation, which is consistent with the destruction of a surface state. We also observe a slight shift towards high-$E_B$ attributed to a shift of the chemical potential at the surface due to the electron doping induced by the K dopant atoms. As discussed below, the states within 3 eV below the Fermi level ($E_F$) are also affected [see Fig. \ref{core}(c)], though in a more complicated way.

Although they cannot be assigned unambiguously, additional peaks in the spectra of the P-substituted samples are observed in the $127\leq E_B \leq 133$ eV range. While some of them may also come from the P $2p$ states, we should expect that others may be related to the Eu $4d$ energy levels. Indeed, as shown in the left inset of Fig. \ref{core}(a), P-free EuFe$_2$As$_2$ exhibits a rather rich spectrum in this region: a series of peaks spaced by an average interval of 920 meV are observed at $E_B=127.54$, 128.48, 129.40, 130.32 and 131.22 eV, in the proximity of a broader bump centered at $E_B=133.8$ eV. Interestingly, similar features have been already reported in Eu metal \cite{Kowalczyk_1974} as well as in EuTe \cite{Shirley_book1978}. Starting from an Eu$^{2+}$ $(4d^{10}4f^7)$ initial state, the spectrum of Eu metal was interpreted in terms of the $^7D$ and $^9D$ spectroscopic terms of the Eu$^{3+}$ $(4d^94f^7)$ final state \cite{Kowalczyk_1974}. As in Eu metal, while the $^9D_{J=(2-6)}$ multiplets of the $^9D$ term can be identified in EuFe$_2$As$_2$, the $^7D$ term remains unresolved.     

\begin{figure}[!t]
\begin{center}
\includegraphics[width=3.4in]{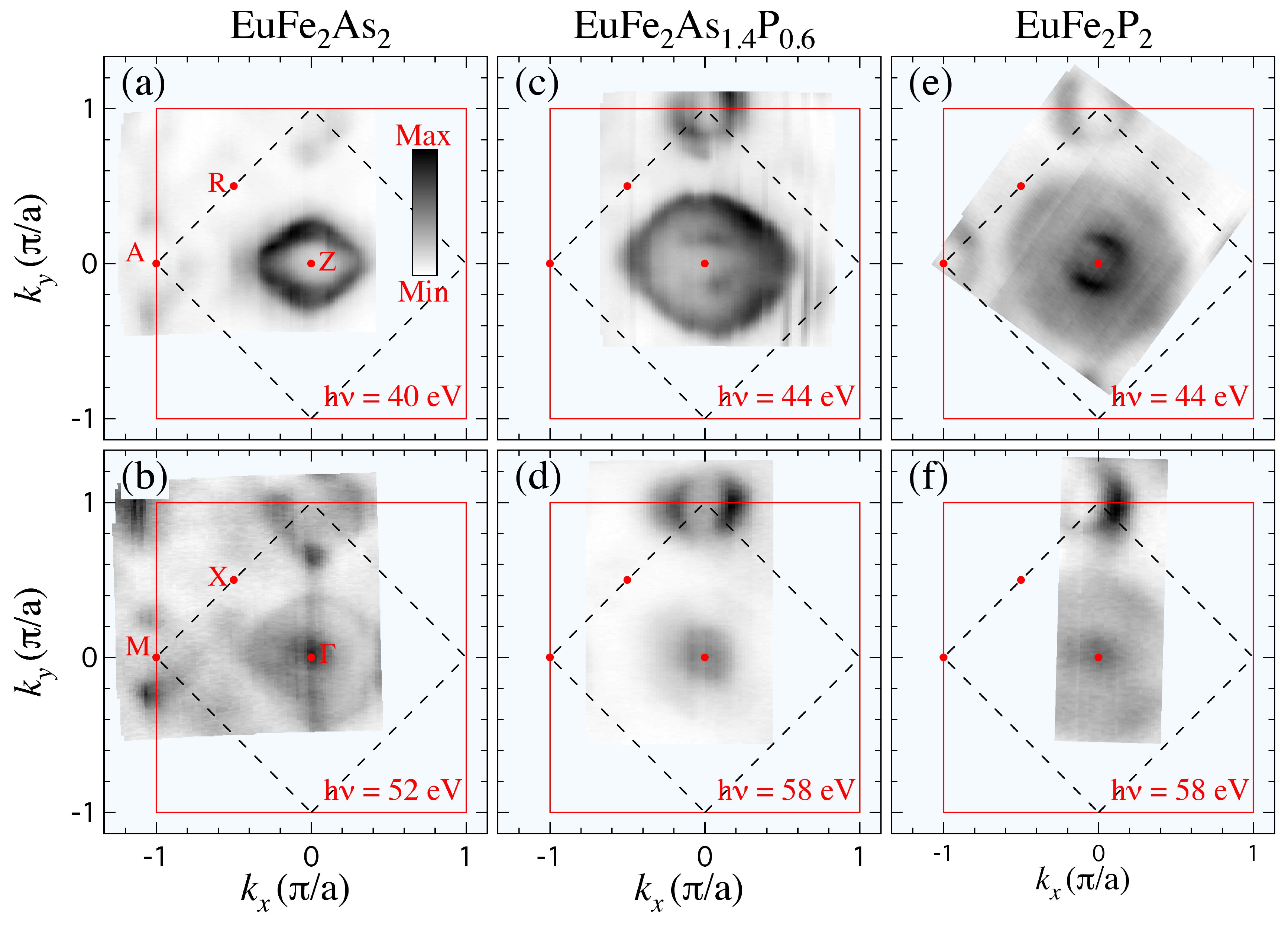}
\end{center}
\caption{\label{FS}(Color online) Fermi surface mappings of EuFe$_2$As$_2$ [(a)-(b)], EuFe$_2$As$_{1.4}$P$_{0.6}$ [(c)-(d)] and EuFe$_2$As$_2$ [(e)-(f)], obtained by integrating the photoemission intensity within $\pm 5$ meV of $E_F$. The top and bottom rows refer to mappings recorded with $k_z\sim \pi/c^{\prime}$ (Z) and $k_z\sim 0$ ($\Gamma$), respectively. The red squares define the in-plane projection of the 1 Fe/unit cell BZ.}
\end{figure}

We now switch our attention to the electronic states forming the Fermi surface (FS) of EuFe$_2$As$_{2-x}$P$_{x}$. In Fig. \ref{FS}, we compare the FSs of EuFe$_2$As$_2$, EuFe$_2$As$_{1.4}$P$_{0.6}$ and EuFe$_2$P$_2$, around $k_z=0$ and $k_z=\pi/c^{\prime}$. As with BaFe$_2$As$_2$, the FS of EuFe$_2$As$_2$ exhibits stronger spots of intensity, mainly visible around the M/A [$(0,\pi,k_z)$] point, which are attributed to the presence of Dirac cones induced by the antiferromagnetic ordering \cite{RichardPRL2010}. As a result of a surface reconstruction similar to the one reported in BaFe$_{2-x}$Co$_x$As$_2$ \cite{vanHeumenPRL106} and Ca$_{0.83}$La$_{0.17}$Fe$_2$As$_2$ \cite{YB_Huang_CPL30}, an extra pattern of intensity is observed at the X point.  

As reported in a previous ARPES study of EuFe$_2$As$_{1.56}$P$_{0.44}$ \cite{Thirupathaiah_PRB84} and commonly expected for a Fe-based superconductor with the 122 crystal structure and a non-magnetically ordered Fe network \cite{RichardRoPP2011}, the FS of EuFe$_2$As$_{1.4}$P$_{0.6}$ and EuFe$_2$P$_2$ are composed by $\Gamma$-centered hole FS pockets and M-centered electron FS pockets. Their size evolves with the P content, but in a non-symmetrical way. While the FS pattern at the M point becomes only a little smaller with $x$ increasing from 0 to 2, the size of the $\Gamma$-centered pockets increases significantly, suggesting a hole-doping that cannot be explained by a simple chemical potential shift, as also pointed out in a previous ARPES study \cite{Thirupathaiah_PRB84}. According to our core level data reported above, we cannot exclude the possibility that this non-trivial doping dependence might be related to a surface doping effect. However, the present case is quite different from the situations encountered for YBa$_2$Cu$_3$O$_{7-x}$ \cite{Nakayama_PRB75,Zabolotnyy_PRB76} and the 1111-ferropnictides \cite{C_LiuPRB2010,Nishi_PRB84}.	In particular, the photoemission intensity mappings displayed in Fig. \ref{FS} for different photon energies indicate non-negligible electronic dispersion along $k_z$, thus suggesting that the low-energy states probed by ARPES cannot be pure two-dimensional surface states.

\begin{figure}[!t]
\begin{center}
\includegraphics[width=3.4in]{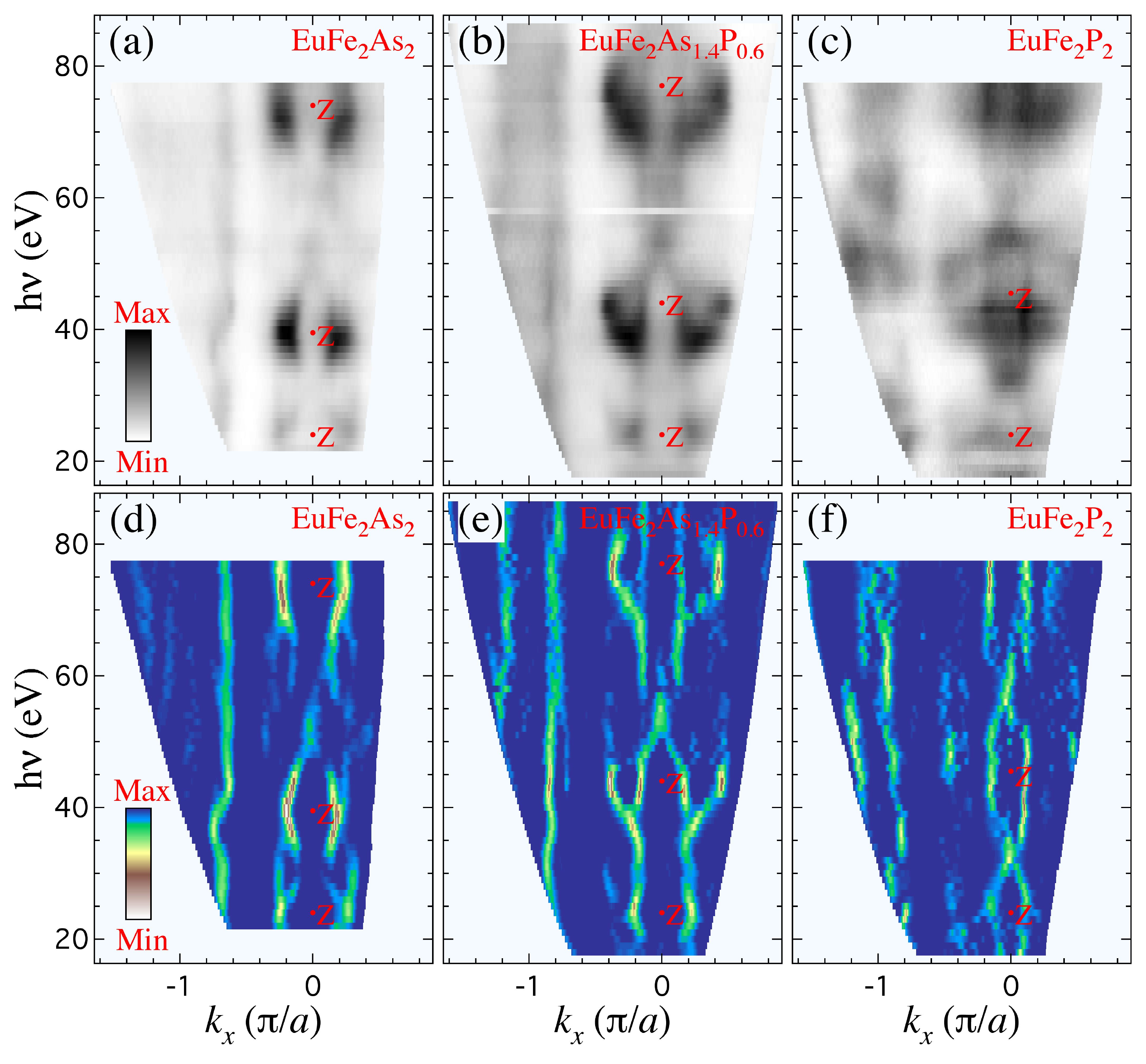}
\end{center}
\caption{\label{kz_map}(Color online) Photon energy dependence of the photoemission intensity at $E_F$ ($\pm 5$ meV integration) along $\Gamma$-M for (a) EuFe$_2$As$_2$, (b) EuFe$_2$As$_{1.4}$P$_{0.6}$ and (c) EuFe$_2$P$_2$, and (d)-(f) their corresponding intensity plots of 1D-curvature (along $k_x$) \cite{P_Zhang_RSI2011}.}
\end{figure}

\begin{figure}[!t]
\begin{center}
\includegraphics[width=3.4in]{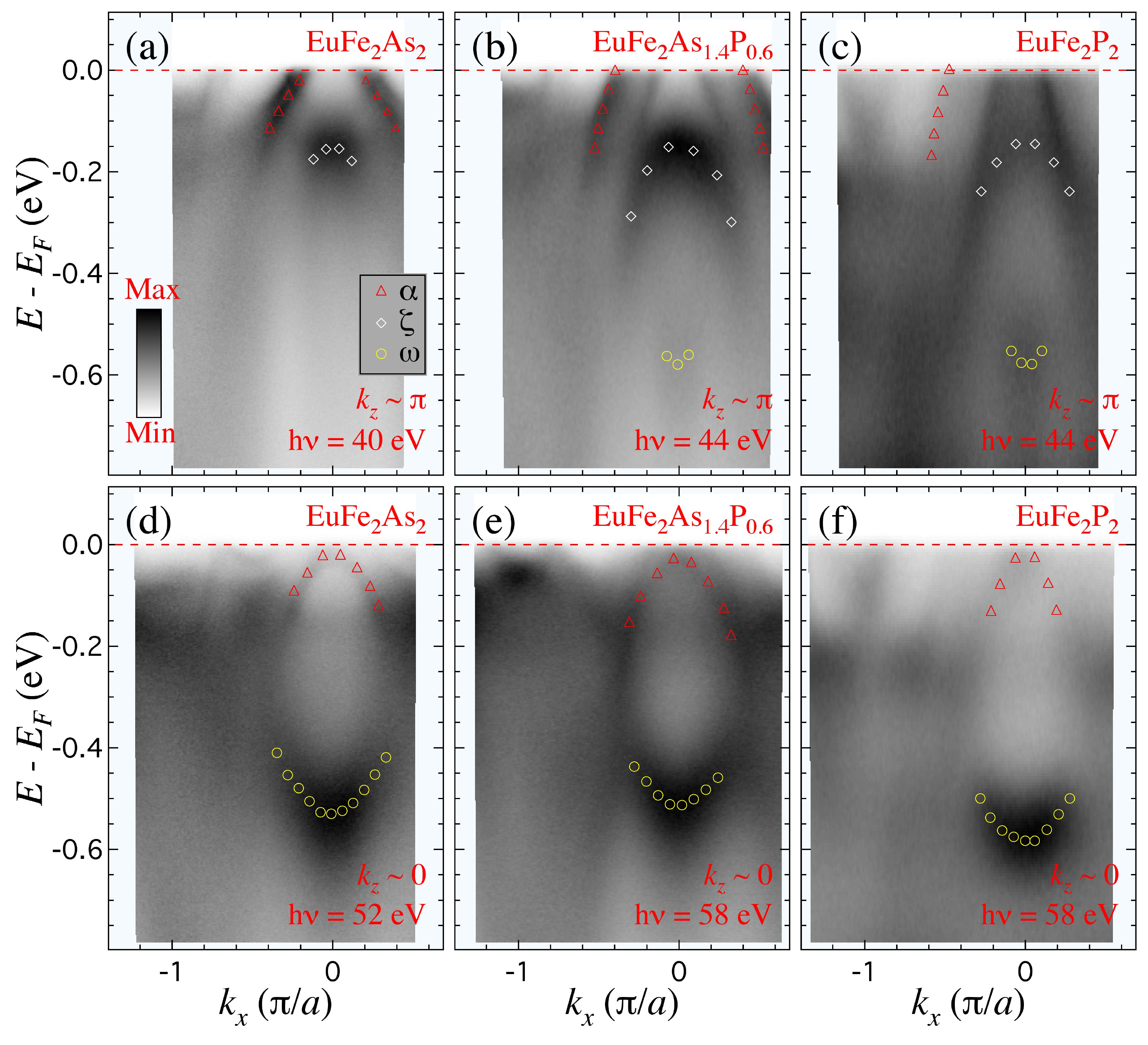}
\end{center}
\caption{\label{dispersion}(Color online) Top row: ARPES intensity plot along the $\Gamma$-M direction for $k_z\sim\pi$. Bottom row: same but for $k_z\sim0$. The open symbols are guides to the eye for the $\alpha$ band with $d_{\textrm{even}}$ orbital character, as well as for the $\zeta$ and $\omega$ bands, which both have a strong $d_{z^2}$ component \cite{Y_Zhang_PRB83}.}
\end{figure}

A better visualization of the electronic dispersion along $k_z$ is provided by the photon energy dependence of the photoemission intensity along the $\Gamma$-M high-symmetry line of the three compounds measured in our study, which are shown in the top row of Fig. \ref{kz_map}, as well as the curvature intensity plots \cite{P_Zhang_RSI2011} given in the bottom row of Fig. \ref{kz_map}. Within the free-electron approximation, there is indeed a monotonic relationship between the incident photon energy and the perpendicular momentum $k_z$ of the photoemitted electrons \cite{DamascelliPScrypta2004} that allows us to interpret these plots as FS mappings in the $k_x-k_z$ plane. One of the bands, centered at $k_z=\pi/c^{\prime}$, exhibits strong dispersion along $k_z$. In fact, the ARPES intensity plots displayed in Fig. \ref{dispersion} indicate that while that band has a very large Fermi wave vector ($k_F$) around $k_z=\pi/c^{\prime}$ (top panels of Fig. \ref{dispersion}), it does not even cross $E_F$ around $k_z =0$ (bottom panels of Fig. \ref{dispersion}) for any of the EuFe$_2$As$_{2-x}$P$_{x}$ materials studied here. In other words, this band forms a 3D hole pocket centered at Z, which becomes larger with the P content increasing, observation consistent with a large 3D FS suggested from de Hass van Alphen measurements in CaFe$_2$P$_2$ \cite{Coldea_PRL103}, although that FS is even larger in the latter case. Interestingly, this band varies similarly to the $\alpha$ band in Ba(Fe$_{1-x}$Ru$_x$)$_2$As$_2$ \cite{Nan_XuPRB86, Brouet_PRL105, Dhaka_prl2012}, which is another nominally non-doped 122-ferropnictide. We thus assign to this band the same $d_{\textrm{even}}$ main orbital character, $d_{\textrm{even}}$ being the even combination of the $d_{xz}$ and $d_{yz}$ orbitals. Our photon energy dependence also indicates that the distance in photon energy between two successive Z points increases, which is consistent with the decrease in the $c$'-axis parameter when As atoms are substituted by smaller P atoms.

\begin{figure}[!t]
\begin{center}
\includegraphics[width=3.4in]{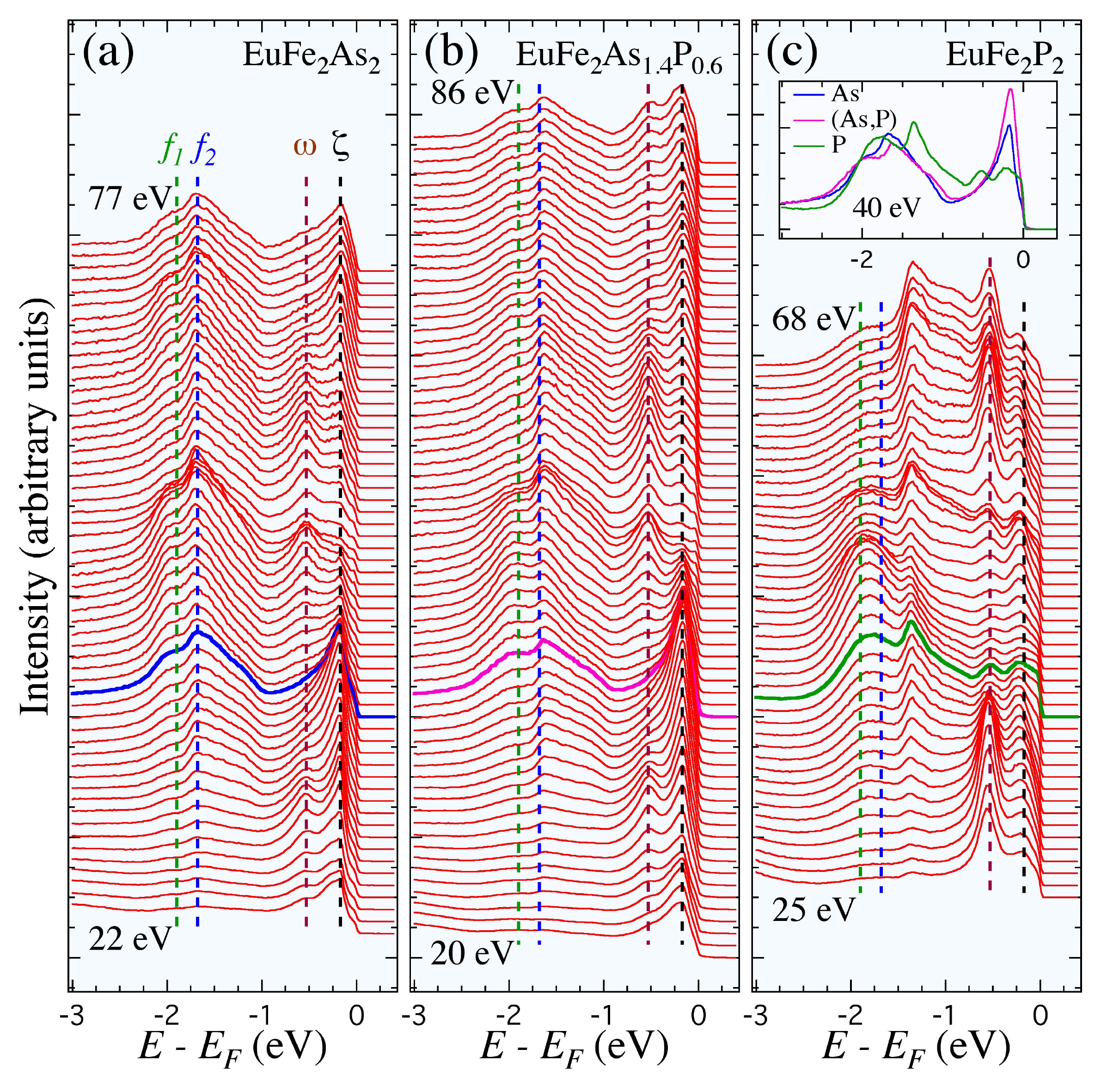}
\end{center}
\caption{\label{Ghv}(Color online) Photon energy dependence of the EDCs recorded at the BZ center in (a) EuFe$_2$As$_2$, (b) EuFe$_2$As$_{1.4}$P$_{0.6}$ and (c) EuFe$_2$P$_2$. The inset in panel (c) compares the EDCs of the 3 compounds recorded with 40 eV photons. As, (As, P) and P refer to EuFe$_2$As$_2$, EuFe$_2$As$_{1.4}$P$_{0.6}$ and EuFe$_2$P$_2$, respectively. In each panel, the vertical lines are guides to the eye for the energy position of 4 peaks detected in EuFe$_2$As$_2$: $f_1$ (-1.9 eV), $f_2$ (-1.7 eV), $\omega$ (-0.53 eV) and $\zeta$ (-0.17 eV).}
\end{figure}

The spectral features shown in Figs. \ref{kz_map} are not as well defined in EuFe$_2$P$_2$ as in the samples with lower P content, an observation further evidenced by the ARPES intensity plots displayed in Fig. \ref{dispersion}. In particular, the near-$E_F$ spectral intensity around the $\Gamma$ point of EuFe$_2$P$_2$ is quite weak [see Fig. \ref{dispersion}(f)]. This contrasts with the intensity of the $\omega$ band found at higher binding energy, which remains high for all $x$ concentrations. In Fig. \ref{Ghv}, we compare the photon energy dependence of the BZ center EDCs of EuFe$_2$As$_2$, EuFe$_2$As$_{1.4}$P$_{0.6}$ and EuFe$_2$P$_2$. The lower energy part of the EDCs is dominated by two peaks coming from the $\zeta$ and $\omega$ bands, which both have a dominant $d_{z^2}$ character \cite{Y_Zhang_PRB83}. Neglecting small $k_z$ and sample composition variations, these peaks are located around 170 and 530 meV below $E_F$, respectively. Interestingly, their intensity oscillates with photon energy (or $k_z$) in anti-phase, as also illustrated in Fig. \ref{dispersion}. While the intensity of the $\zeta$ peak is the strongest around $k_z=\pi/c^{\prime}$ and the weakest around $k_z=0$, the opposite behavior is found for the $\omega$ band. Although our study does not allow us to identify unambiguously the origin of the partial suppression of spectral intensity for the electronic states at low energy, it is possibly related to the surface state identified from the core level data presented in Fig. \ref{core}, which show significantly larger core level splittings in EuFe$_2$P$_2$. Distortion at the surface could eventually introduce additional scattering that would suppressed the coherence of the low-energy states. We caution though that an old study of structural characterization revealed the presence of FeP or Fe$_2$P as secondary phase in the growth of some LnFe$_2$P$_2$ (Ln = lanthanide) compounds \cite{Reehuis_JPCS51}, which could also alter the coherence of the low-energy states. However, we did not detect evidence for these impurities in our samples. 

In addition to the $\zeta$ and $\omega$ peaks, additional spectral intensity is found between 1 and 2.5 eV below $E_F$. In particular, a peak labeled $f_2$ and a shoulder labeled $f_1$ are detected in EuFe$_2$As$_2$ at -1.7 eV and -1.9 eV, respectively. These dispersionless features are not observed in the more commonly studied Ba(Fe$_{1-x}$Co$_x$)$_{2-x}$As$_2$ and Ba$_{1-x}$K$_x$Fe$_2$As$_2$ compounds \cite{Neupane_PRB2011}. In agreement with a previous ARPES study \cite{Adhikary_JPCM25}, we ascribe them to Eu $^4f$ electronic states. As indicated by the inset of Fig. \ref{Ghv}(c), which compares the EDCs of the 3 measured compounds recorded with 40 eV photons, only small changes occur when the P content varies from $x=0$ to 0.6. While the $f_1$ peak position in EuFe$_2$As$_{1.4}$P$_{0.6}$ is barely changed, the $f_2$ peak moves by only about 50 meV towards $E_F$. As illustrated in Fig. \ref{Ghv}(c), the $f_2$ peak shift is much larger in EuFe$_2$P$_2$. As compared to EuFe$_2$As$_2$, the $f_1$ and $f_2$ peaks in EuFe$_2$P$_2$ are located 80 meV and 320 meV closer to $E_F$, respectively. 

Such large splitting between the $f_1$ and $f_2$ peaks allows us to distinguish their spectral lineshapes, which are quite different. In contrast to the $f_1$ peak, which is quite broad and rounded, the $f_2$ peak is rather sharp. The latter is also asymmetric, with a tail on the low $E_B$ side possibly due to the presence of additional peaks, as mainly suggested from the spectra recorded at the highest photon energies. Interestingly, as shown in Fig. \ref{core}(c), our investigation of the electronic states in a EuFe$_2$As$_2$ sample upon K evaporation on the surface indicates that while the shape and intensity of the $f_2$ peak are barely modified by the evaporation of K atoms, the intensity of the $f_1$ peak is reduced, suggesting a surface state. However, this particular observation does not explain the large shift in the position of the $f_2$ peak in EuFe$_2$P$_2$. It is important to note that we did not observe any significant modification of the spectral lineshape across the Eu$^{2+}$ magnetic ordering transition, thus suggesting that it is not relevant for this particular lineshape nor for the large $f_1$-$f_2$ splitting in EuFe$_2$P$_2$. However, the inset of Fig. \ref{Ghv}(c) seems to show that the shift of the $f_2$ peak in EuFe$_2$P$_2$ is accompanied by a spectral weight transfer from the near-$E_F$ states to the $0.5\leq E_B\leq 1.5$ energy range. Whether this could be caused by enhanced Fe-Eu interactions is not excluded but would require confirmation from further theoretical investigations.  

From the core levels to the electronic states in the vicinity of $E_F$, our results indicate the presence of a surface state effect in EuFe$_2$As$_{2-x}$P$_{x}$ that does exist, or at least that does not manifest itself significantly, in the commonly studied Ba(Fe$_{1-x}$Co$_x$)$_{2-x}$As$_2$ and Ba$_{1-x}$K$_x$Fe$_2$As$_2$ 122-ferropnictides. We note that the size of the Eu$^{2+}$ ion is significantly smaller than that of Ba$^{2+}$, and the substitution of P by As contributes to reduce the $c$-axis length even further. Previous studies showed that when the $c$-axis becomes small and the As-As separation in the 122-ferropnictides becomes smaller than about 3 \AA, either due to rare earth substitution \cite{Saha_PRB2012} or As$\rightarrow$P substitution in CaFe$_2$As$_{2-x}$P$_x$ \cite{HL_Shi_JPCM22,KasaharaPRB83}, interactions between successive As layers induce a further decrease in the $c$-axis length of these materials that thus encounter a tetragonal-collapsed tetragonal transition \cite{Saha_PRB2012}, as also supported by theoretical predictions \cite{YildirimPRL102}. We point out that any material in the collapsed tetragonal phase or at the proximity of this transition may show a discontinuity in the pnictide-pnictide interactions at the surface, possibly altering the surface electronic properties. Indeed, a previous ARPES study revealed anomalies in conventional ARPES data recorded on LaRu$_2$P$_2$ as compared to bulk-sensitive soft-x-ray ARPES data \cite{Razzoli_PRL108}. We predict that other 122 materials with similarly small $c$-axis as EuFe$_2$As$_{2-x}$P$_x$ and LaRu$_2$P$_2$ may exhibit the same kind of anomalies reported here and in ref. \cite{Razzoli_PRL108}. Yet, further theoretical and experimental investigations are necessary to confirm or infirm this conjecture.

In summary, we performed an ARPES study of EuFe$_2$As$_{2-x}$P$_{x}$ that shows the evolution of the electronic structure upon increasing the P content. Anomalies in the core level and near-$E_F$ state data suggest the existence of a surface state. Nevertheless, strong $k_z$ modulations enhanced with P substitution are observed in all studied samples for at least one band. Finally, a sudden and unexplained jump in the energy position of one peak associated to the Eu$^{2+}$ $f$ states is observed in EuFe$_2$P$_2$.

We acknowledge M. Shi, E. Razolli and J.-X. Yin for useful discussions. This work was supported by grants from CAS (2010Y1JB6), MOST (2010CB923000,  2011CBA001000, 2011CBA00102, 2012CB821403 and 2013CB921703) and NSFC (10974175, 11004232, 11034011/A0402 and 11274362) from China. This work is based in part on research conducted at the Synchrotron Radiation Center, which is primarily funded by the University of Wisconsin-Madison with supplemental support from facility Users and the University of Wisconsin-Milwaukee. The Advanced Light Source is supported by the Director, Office of Science, Office of Basic Energy Sciences, of the U.S. Department of Energy under Contract No. DE-AC02-05CH112. The work at ORNL was supported by the Department of Energy, Basic Energy Sciences, Materials Sciences and Engineering Division.

\bibliography{biblio_long}

\end{document}